\documentclass{emulateapj}
\usepackage{amsmath}
\usepackage{graphicx}
\shorttitle{Nonlinear mirror and whistler instabilities}
\shortauthors{}

\begin{document}
%\title{PIC simulations of electron anisotropy evolution in collisionless astrophysical plasmas: the role of the mirror and whistler instabilities}
\title{PIC Simulations of the Effect of Velocity Space Instabilities on Electron Viscosity and Thermal Conduction}  

%-----
%\author{Authors}
\author{Mario A. Riquelme\altaffilmark{1}, Eliot Quataert\altaffilmark{2} \& Daniel Verscharen\altaffilmark{3}}
\altaffiltext{1}{Departamento de F\'isica, Facultad de Ciencias F\'isicas y Matem\'aticas, Universidad de Chile; mario.riquelme@dfi.uchile.cl}
\altaffiltext{2}{Astronomy Department and Theoretical Astrophysics Center, University of California, Berkeley, CA 94720; eliot@berkeley.edu}
\altaffiltext{3}{Space Science Center and Department of Physics, University of New Hampshire, Durham, NH 03824; Daniel.Verscharen@unh.edu}
%-----

\begin{abstract} 

\noindent In low-collisionality plasmas, velocity-space instabilities are a key mechanism providing an effective collisionality for the plasma.  We use particle-in-cell (PIC) simulations to study the interplay between electron- and ion-scale velocity-space instabilities and their effect on electron pressure anisotropy, viscous heating, and thermal conduction.  The adiabatic invariance of the magnetic moment in low-collisionality plasmas leads to pressure anisotropy, $\Delta p_j \equiv p_{\perp,j} - p_{||,j} > 0$, if the magnetic field \textbf{\textit{B}} is amplified ($p_{\perp,j}$ and $p_{||,j}$ denote the pressure of species $j$ [electron, ion] perpendicular and parallel to \textbf{\textit{B}}). If the resulting anisotropy is large enough, it can in turn trigger small-scale plasma instabilities. Our PIC simulations explore the nonlinear regime of the mirror, ion-cyclotron, and electron whistler instabilities, through continuous amplification of the magnetic field $|\textbf{\textit{B}}|$ by an imposed shear in the plasma. In the regime $1 \lesssim \beta_j \lesssim 20$ ($\beta_j \equiv 8\pi p_j/|\textbf{\textit{B}}|^2$), the saturated electron pressure anisotropy, $\Delta p_e/p_{||,e}$, is determined mainly by the (electron-lengthscale) whistler marginal stability condition, with a modest factor of $\sim 1.5-2$ decrease due to the trapping of electrons into ion-lengthscale mirrors.  We explicitly calculate the mean free path of the electrons and ions along the mean magnetic field and provide a simple physical prescription for the mean free path and thermal conductivity in low-collisionality $\beta_j \gtrsim 1$ plasmas.  Our results imply that velocity-space instabilities likely decrease the thermal conductivity of plasma in the outer parts of massive, hot, galaxy clusters.  We also discuss the implications of our results for electron heating and thermal conduction in low-collisionality accretion flows onto black holes, including Sgr A* in the Galactic Center.\newline \end{abstract}
% even when the timescale for field amplification is $\sim 10^3$ times the ion-cyclotron period.  
%Violation of magnetic moment conservation allows the plasma to maintain marginal stability to the mirror instability. 
%$|\delta \vec{B}| \sim 0.1 |<\vec{B}>|$, 
%the secular stage is accompanied by significant violation of 
%The fluctuation amplitude at saturation is $|\delta \vec{B} \sim 0.3 |<\vec{B}>|$.
%Our setup differs significantly from most studies of ion velocity space instabilities, which focus on the initial value problem of a finite temperature anisotropy that relaxes to marginal stability.  

%\keywords{kinetic plasma effects -- solar wind}

\section{Introduction}
\label{sec:intro}

\noindent In a low collisionality plasma, differences in pressure along ($p_{\parallel,j}$) and perpendicular ($p_{\perp,j}$) to the local magnetic field $\textbf{\textit{B}}$ are produced by compression, shearing, and/or heating of the plasma.  This is a consequence of the adiabatic invariance of the `bounce invariant' and the magnetic moment, $\mu_j \equiv v_{\perp,j}^2/B$, where $v_{\perp,j}$ is the velocity perpendicular to the local magnetic field, $B=|\textbf{\textit{B}}|$ and $j$ denotes the particle species \citep{Kulsrud1983}.  Thus, absent Coulomb collisions, magnetic field amplification and/or plasma compression generically drive $p_{\perp,j} > p_{||,j}$, while a decrease in the magnetic field strength and/or plasma expansion generically drives $p_{\perp,j} < p_{||,j}$.  These pressure anisotropies can be dynamically important because they modify the effective magnetic tension in the plasma, produce an effective viscosity, and can drive velocity-space instabilities.  Systems where $p_{\perp_j} \ne p_{||_j}$ is believed to be important are low-luminosity accretion flows around compact objects \citep{SharmaEtAl07}, the intracluster medium (ICM) \citep{SchekochihinEtAl05, Lyutikov07}, and the heliosphere \citep[][]{MarucaEtAl11,RemyaEtAl13}. \newline
%the electron pressures parallel and perpendicular to the magnetic field ($p_{||,e}$ and $p_{\perp,e}$, respectively) are expected to evolve differently, making $\Delta p_e \equiv p_{\perp,e} - p_{||,e} \ne 0$. Examples of systems where electron pressure anisotropies are important are low-luminosity accretion flows around compact objects \citep{SharmaEtAl07}, the intracluster medium (ICM) \citep{SchekochihinEtAl05, Lyutikov07}, and the heliosphere \citep[][]{MarucaEtAl11,RemyaEtAl13}. \newline

\noindent When $p_{\perp,e} > p_{||,e}$, the electron whistler instability is excited and can limit the amount of electron pressure anisotropy that develops \citep{GaryEtAl96}.
%If magnetic field decreases, the expectation is that on average $p_{\perp,e} < p_{||,e}$, making the electron firehose instability the main candidate to limit the electron anisotropy [refs]. 
Although the linear behavior of the whistler instability is well understood, the long term, nonlinear evolution of the instability is less clear.  Indeed, in many astrophysically relevant cases, the generation of pressure anisotropy occurs over time scales longer than the initial exponential growth phase that characterizes velocity-space instabilities. Moreover, the ions are expected to develop their own pressure anisotropy in synch with the electrons, giving rise to analogous instabilities. For example, when $B$ grows, the mirror and ion-cyclotron (IC) instabilities regulate the ion pressure anisotropy \citep{Hasegawa69, Gary92,Southwood93}.\newline

\noindent \cite{Kunz2014} (using hybrid PIC) and \cite{RiquelmeEtAl2015} (using full PIC with ion to electron mass ratios $m_i/m_e=1-10$) studied the saturation of ion velocity-space instabilities in a model problem where background velocity shear amplifies/reduces the strength of a background magnetic field, thus continually driving pressure anisotropy. \cite{HellingerEtAl08} and \cite{SironiEtAl2015a} used an expanding/compressing box to accomplish the same goal of continually driving pressure anisotropy.  Continuous driving of the pressure anisotropy is important because it allows one to study the nonlinear saturation of velocity-space instabilities, in contrast to more standard initial value calculations.  One of the conclusions of the works of \cite{Kunz2014} and \cite{RiquelmeEtAl2015} is that the mirror instability reaches nonlinear amplitudes with $\delta B \sim B$, independent of how slowly the magnetic field is amplified relative to the ion-cyclotron period ($\delta B \equiv |\delta \textbf{\textit{B}}|$ and $\delta \textbf{\textit{B}} \equiv \textbf{\textit{B}} - <\textbf{\textit{B}}>$; throughout this paper, $< >$ will stand for the average of a quantity over the simulation volume or over a population of particles, depending on the quantity under consideration).  A natural question is what effect these large amplitude mirrors have on the electron physics and how the whistler instability grows in the presence of large amplitude mirror modes.\newline
% In \cite  {RiquelmeEtAl2015  } we performed a particle-in-cell (PIC) simulation study of the nonlinear evolution of the mirror and IC instabilies, using small values of the ion to electron mass ratio $m_i/m_e=1-10$. We found that the mirror modes dominate for $\beta_i \gtrsim 1$ ($\beta_i \equiv 8\pi p_i/B^2$, where $p_i$ is the ion pressure), with the IC modes playing a gradually less important role as $\beta_i$ grows.  \citep{Kunz2014} used the hybrid PIC method in a related study of the saturation of velocity space instabilities at high $\beta_i \sim 200$ and found that in this regime the mirror instability completely dominates over the ion-cyclotron instability.  

\noindent In this paper we use particle-in-cell (PIC) simulations to study the combined effect of mirror and whistler instabilities on the electron pressure anisotropy.  This work is thus an extension of our previous study of the ion-scale, kinetic instabilities \citep{RiquelmeEtAl2015}. In order to properly separate phenomena occurring on the ion- and electron-lengthscales, in this work we use larger values of the ion to electron mass ratio: $m_i/m_e=$ 64 and 128.  We focus throughout this paper on a fiducial case with initial $\beta_i=\beta_e=20$ (in our simulations, $\beta_j$ decreases in time as the background magnetic field is amplified). At sufficiently low $\beta_i \lesssim 1$, the ion-cyclotron instability is expected to be more important than the mirror instability in regulating the ion pressure anisotropy.  However, \citet{RiquelmeEtAl2015} did not find any significant differences in the nonlinear mirror evolution for $\beta_i=20$ and $\beta_i = 80$, and even at $\beta_i = 6$ the ion-cyclotron instability was sub-dominant.  Moreover, \citet{Kunz2014} found similar results at higher $\beta_i = 200$.  Thus we believe that the calculations presented in this paper with $\beta_i=20$ provide a good model for the saturation of electron velocity-space instabilities in $\beta_i \gtrsim 1$ plasmas (at least for similar ion and electron temperatures, $T_i \sim T_e$; see \S \ref{sec:conclu}).\newline

\noindent Our work has two important applications. First, the nonlinear evolution of the mirror and whistler instabilities effectively sets the pitch-angle scattering rate (and thus the mean free path) of electrons in low collisionality plasmas.  This in turn determines the viscosity and thermal conductivity of the plasma.  In our calculations we will directly measure the electron mean free path $\langle \lambda_e \rangle$ in the collisionless regime, and infer its physical dependence on the plasma parameters.  Secondly, electron pressure anisotropy generates an ``anisotropic viscosity" that can contribute to the heating of electrons in accretion disks and other low collisionality plasmas \citep{SharmaEtAl07}.  We will see that the magnitude of this heating depends on the magnitude of the electron pressure anisotropy produced by the combined effect of the whistler and mirror instabilities.  %And second, an important component in determining the temperature profile of collisionless systems is their ability to transport heat, expected to depend on the mean free path of the electrons along the background magnetic field, $\langle \lambda \rangle_e$. The nonlinear behavior of the mirror and whistler instabilities (which will provides us with effective scattering rates for both species) will allow us to measure $\langle \lambda \rangle_e$ directly, and to determine its physical dependence on the plasma parameters.
\newline

This paper is organized as follows. In \S \ref{sec:numsetup} we describe the numerical set up of our runs, and our simulation strategy. In \S \ref{sec:interplay} we determine the saturated pressure anisotropy $\Delta p_e$ for the electrons and quantify the electron heating due to anisotropic viscosity. In \S \ref{sec:conduct} we measure the mean free path of electrons and ions, and determine their dependence on the physical parameters of the plasma.  In \S \ref{sec:conclu} we summarize our results and discuss their implications for galaxy clusters and low-collisionality black hole accretion flows.

\begin{figure*}[t!]  \centering \includegraphics[width=18cm]{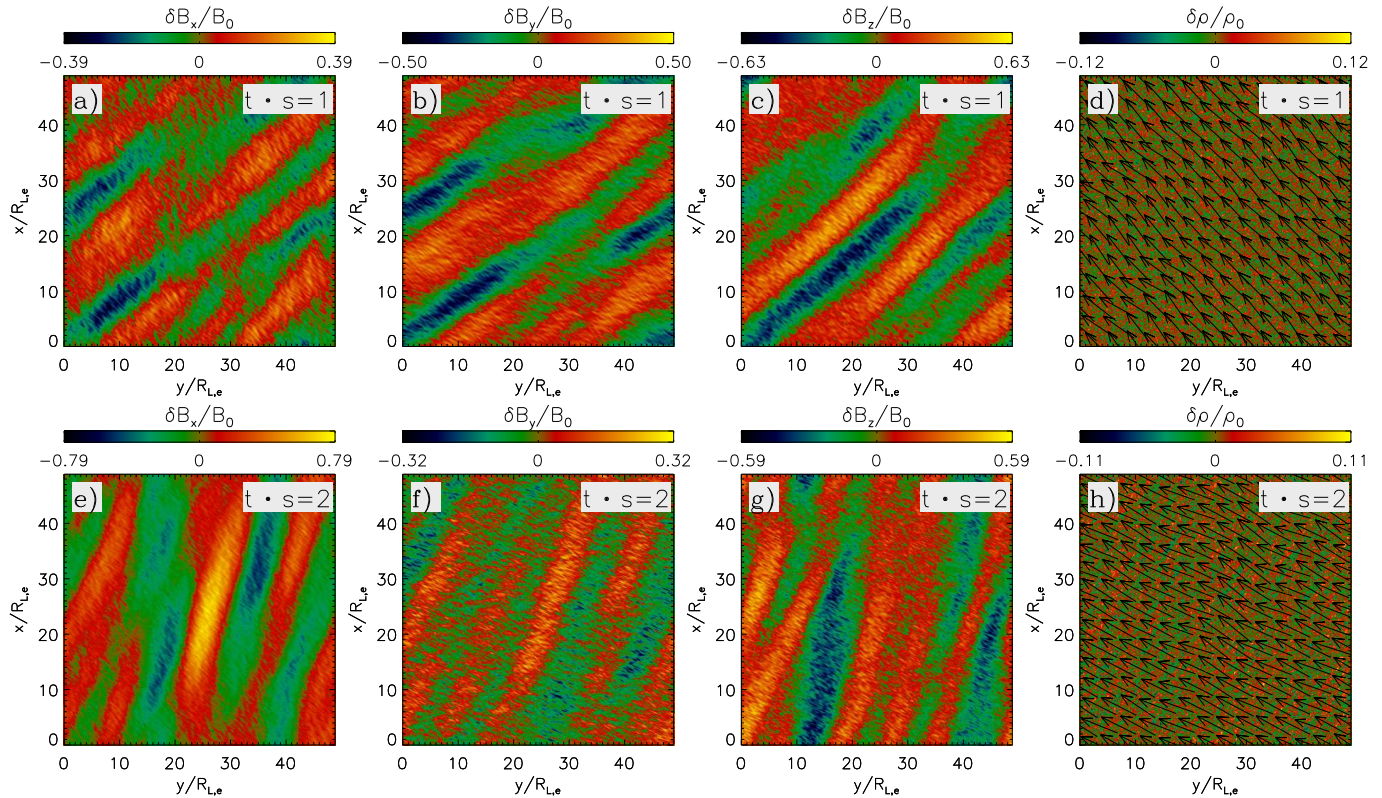}%{beta6.ps}
  \caption{The three components of $\delta \textbf{\textit{B}}$ and plasma density fluctuations $\delta \rho$ at two different times: $t\cdot s = 1$ (upper row) and $t\cdot s = 2$ (lower row), for a simulation with only whistlers modes (run OW1 with $m_i/m_e = \infty$, so that the ions only provide a neutralizing charge). Fields and density are normalized by the initial magnetic field and density, $B_0$ and $\rho_0$, respectively. Arrows in panels $d$ and $h$ show the mean magnetic field direction on the $x-y$ simulation plane. For this $m_i/m_e = \infty$ case, the magnetic fluctuations are dominated by nearly parallel whistler modes.}
\label{fig:fieldsbeta6}
\end{figure*}

\section{Simulation Setup}
 \label{sec:numsetup}
We use the electromagnetic, relativistic PIC code TRISTAN-MP \citep{Buneman93, Spitkovsky05} in two dimensions. The simulation box consists of a square box in the $x-y$ plane, containing plasma with a homogeneous initial magnetic field $\textbf{\textit{B}}_0=B_0 \hat{x}$. Since we want to simulate a magnetic field that is being amplified in an incompressible way, we impose a velocity shear so that the mean particle velocity is $\textbf{\textit{v}} = -sx\hat{y}$, where $s$ is a shear parameter with units of frequency and $x$ is the distance along $\hat{x}$. From flux conservation, the $y$-component of the mean field evolves as $d \langle B_y\rangle /dt = -sB_0$. This implies a net growth of $|\langle \textbf{\textit{B}}\rangle |$, which in turn drives $p_{\perp,j} > p_{||,j}$ during the whole simulation. \newline
 
Simulations resolving the $x-y$ plane can capture mirror, IC, and whistler modes with wave vectors $\textbf{\textit{k}}$ forming any angle with the mean magnetic field $\langle  \textbf{\textit{B}}\rangle$. In our previous study, focused on the interplay between the mirror and IC instabilities only \citep{RiquelmeEtAl2015}, we found that most of the physics of the relevant instabilities is captured when the $x-y$ plane is resolved (as in the simulations in this paper).\newline 

\noindent The key parameters in our simulations are the particles' magnetization, quantified by the ratio between the initial cyclotron frequency of each species and  the shear rate of the plasma, $\omega_{c,j}/s$ ($j=i,e$), and the ion to electron mass ratio, $m_i/m_e$.  In typical astrophysical environments, $\omega_{c,j} \gg s$. Due to computational constraints, we will use values of $\omega_{c,j}/s \gg 1$, but still much smaller than expected in real astrophysical settings. Because of this, we have made sure to reach the regime where both $m_i/m_e$ and $\omega_{c,j}/s$ are large enough so that their values do not qualitatively affect any of our conclusions.\newline

\noindent Our simulations have initial $\beta_i = \beta_e = 20$. In all of our runs $k_BT_e/m_ec^2 = 0.28$, which implies $\omega_{c,e}/\omega_{p,e}=0.17$ (where $k_B$, $T_e$, and $\omega_{p,e}$ are the Boltzmann's constant, the electron temperature, and the electron plasma frequency). Thus the varying physical parameters in our simulations will be: $\omega_{c,e}/s$ and $m_i/m_e$ (which uniquely fix $\omega_{c,i}/s$ and $k_{B}T_i/m_ic^2$). Some of our simulations use ``infinite mass ions" (the ions are technically immobile, so they just provide a neutralizing charge), with the goal of focusing on the electron-scale physics.   These provide a useful contrast with our finite $m_i/m_e$ runs and allow us to isolate the impact of ion physics on the electrons. The numerical parameters in our simulations will be: N$_{\textrm{ppc}}$ (number of particles per cell), $c/\omega_{p,e}/\Delta_x$ (the electron skin depth in terms of grid size), $L/R_{L,i}$ (box size in terms of the initial ion Larmor radius for runs with finite $m_i/m_e$; $R_{L,i} = v_{th,i}/\omega_{c,i}$, where $v_{th,i}^2=k_BT_i/m_i$ is the rms ion velocity), and $L/R_{L,e}$ (box size in terms of the initial electron Larmor radius for runs with infinite $m_i/m_e$). Table \ref{table:1D} shows a summary of our key simulations. We ran a series of simulations ensuring that the numerical parameters (e.g., different N$_{\textrm{ppc}}$) do not significantly affect our results. Note that most runs used just for numerical convergence are not in Table \ref{table:1D}.\newline
%\footnote{Numerical convergence was check in all of the runs in Table \ref{table:1D}, so they are a subset of all the runs used in the study.}
%Thus, the physical parameters in our runs will be: $\omega_{c,e}/s$ and $m_i/m_e$; and the numerical parameters will be: N$_{\textrm{ppc}}$ (number of particles per cell), $c/\omega_{p,e}/\Delta_x$ (the electron skin depth in terms of the grid points separation), $L/R_{L,i}$ (box size in terms of the ion Larmor radius for runs with finite $m_i/m_e$), and $L/R_{L,e}$ (box size in terms of the electron Larmor radius for runs with infinite $m_i/m_e$). As mention above, we ran a series of simulations making sure that neither the numerical parameters or the physical parameters $\omega_{c,e}/s$ and $m_i/m_e$ do not play any role in our results. However, we list in Table 1 only those simulations used to present our results.  
\begin{deluxetable}{lllllll} \tablecaption{Physical and numerical parameters of the simulations} \tablehead{ \colhead{Runs}&\colhead{$m_i/m_e$}&\colhead{$\omega_{c,e}/s$}&\colhead{$c/\omega_{p,e}/\Delta_x$}&\colhead{N$_{\textrm{ppc}}$}&\colhead{L/R$_{L,i}$}&\colhead{L/R$_{L,e}$} } \startdata
  OW1 &  $\infty$& 2500 &  5 & 60 & - & 48\\
  OW2 &  $\infty$& 1000 &  5 & 20 & - & 48 \\
  %  OW3 &  $\infty$& 5000 &  5 & 60 & - & 25 \\
  MW1 &  64& 2500 &  5 & 20 & 22 & 176 \\
  MW2 &  64& 2500 &  5 & 60 & 22 & 176 \\
  MW3 & 128& 5000 & 5 & 40 & 22 & 242 \enddata \tablecomments{A summary of the physical and numerical parameters of the simulations discussed in the paper. These are the mass ratio $m_i/m_e$, the initial electron magnetization $\omega_{c,e}/s$, the electron skin depth $c/\omega_{p,e}/\Delta_x$ (where $\Delta_x$ is the grid point separation), the number of particles per cell N$_{\textrm{ppc}}$ (including ions and electrons), the box size in units of the typical initial ion Larmor radius $L/R_{L,i}$ ($R_{L,i} = v_{th,i}/\omega_{c,i}$, where $v_{th,i}^2=k_BT_i/m_i$ is the rms ion velocity and $k_B$ and $T_i$ are the Boltmann constant and the ion temperature, respectively), and the box size in terms of the typical initial electron Larmor radius $L/R_{L,e}$. We confirmed numerical convergence by varying resolution in $c/\omega_{c,e}/\Delta_x$, N$_{ppc}$, and L/R$_{L,i}$ (L/R$_{L,e}$). All of the runs have $\beta_i=\beta_e=20$, $kT_e/m_ec^2=0.28$, $\omega_{c,e}/\omega_{p,e}=0.17$, and $c=0.225 \Delta_x/\Delta_t$, where $\Delta_t$ is the simulation time step.} \label{table:1D} \end{deluxetable} 
 
\section{Electron Physics in Collisionless Shearing Flows}
\label{sec:interplay}

In this section we quantify the nonlinear evolution of the electron pressure anisotropy in collisionless shearing flows, taking into account the combined effect of the nonlinear whistler and mirror instabilities. In order to understand the relative importance of the whistler and mirror modes, we start by isolating the effect of the whistler instability. We do this by using simulations where the ion mass is set to infinity, so that only the electron-scale whistler instability can grow.
\begin{figure}%[t!]
\centering
\includegraphics[width=9cm]{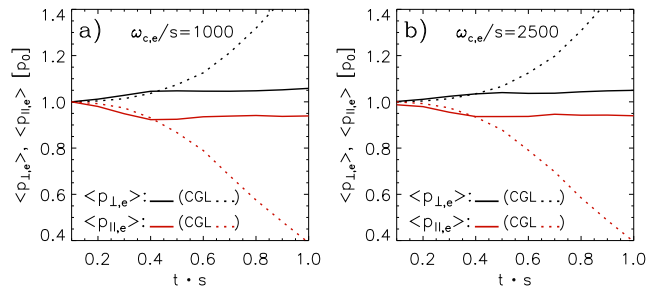}%{beta6.ps}
\caption{The initial evolution of the electron pressures perpendicular (black-solid) and parallel (red-solid) to $\textbf{\textit{B}}$ for runs OW2 and OW1 in Table \ref{table:1D} (with $\omega_{c,e}/s=1000$ and $\omega_{c,e}/s=2500$, respectively). The black- and red-dotted lines show the expectation for the perpendicular and parallel pressures from the CGL or double adiabatic limit \citep{ChewEtAl56}. Significant deviation from adiabatic evolution can be seen at $t \cdot s \gtrsim 0.4$.}
\label{fig:cgl_1000and2500}
\end{figure}
\subsection{Simulations with Whistlers Only}
\label{sec:whistlersonly}
Figure \ref{fig:fieldsbeta6} shows the magnetic field fluctuations and plasma density in a simulation in which the ions have an infinite mass and thus there is no mirror or IC instabilities (run OW1 in Table \ref{table:1D}).  In this simulation, the electrons are only affected by the whistler instability that develops when the electron pressure anisotropy increases due to the growth of the background magnetic field. The upper row in Figure \ref{fig:fieldsbeta6} is at $t\cdot s=1$, i.e., after one shear time, while the lower row is at $t\cdot s=2$.   Figure \ref{fig:fieldsbeta6} shows that at all times the magnetic fluctuations are dominated by the nearly parallel whistler modes, with wavenumbers $k$ satisfying $kR_{L,e} \sim 0.5$; there are no ion-lengthscale fluctuations contributing to $\delta \textbf{\textit{B}}$. 
\begin{figure}[t!]  
\centering 
\includegraphics[width=9cm]{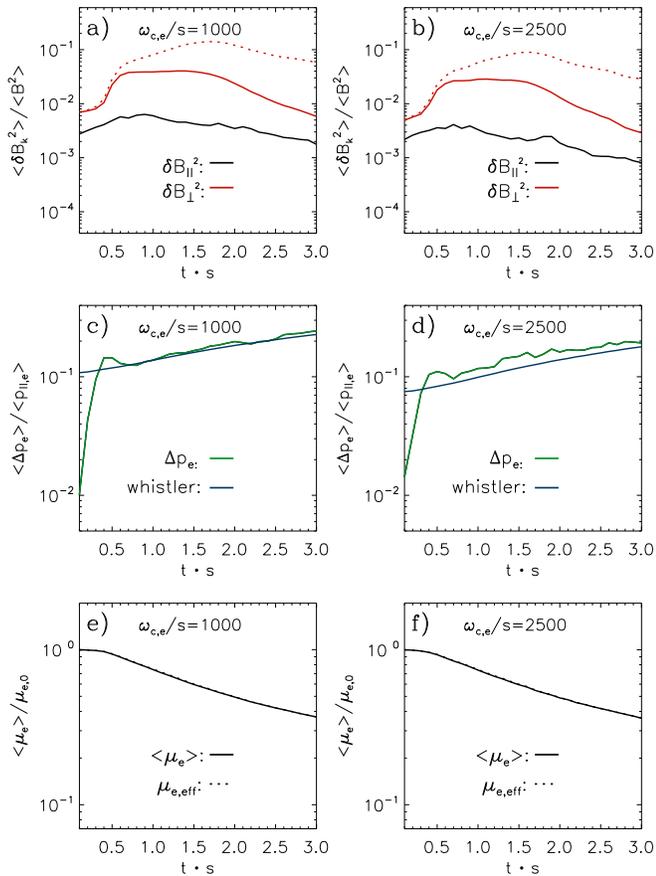}%{beta6.ps}
\caption{The evolution of different volume-averaged quantities for two simulations with $\omega_{c,e}/s=1000$ (run OW2; left column) and $\omega_{c,e}/s=2500$ (run OW1; right column), which use $m_i/m_e=\infty$ (the ions simply provide a neutralizing charge). {\it Upper row:} the volume-averaged magnetic energy components parallel and perpendicular to $\langle \textbf{\textit{B}}\rangle$, $\delta B_{||}^2$ (black) and $\delta B_{\perp}^2$ (red), respectively (normalized by $\langle B^2\rangle$). For comparison, the red-dotted lines show $\delta B_{\perp}^2$ normalized by $B_0^2$. {\it Middle row:} the electron pressure anisotropy (green line), with the linear whistler instability thresholds for growth rates $\gamma_w = 5s$ (black line). The pressure anisotropy saturates at a value consistent with the linear instability threshold of the $\gamma_w = 5s$ modes. {\it Lower row:} the electron magnetic moment; see equation \ref{eq:mu} and associated discussion for definitions of $\mu_e$ (solid) and $\mu_{e,eff}$ (dotted).} 
\label{fig:benandanis_wces1000and5000} 
\end{figure} 
\begin{figure*}[t!]
\centering
  \includegraphics[width=18cm]{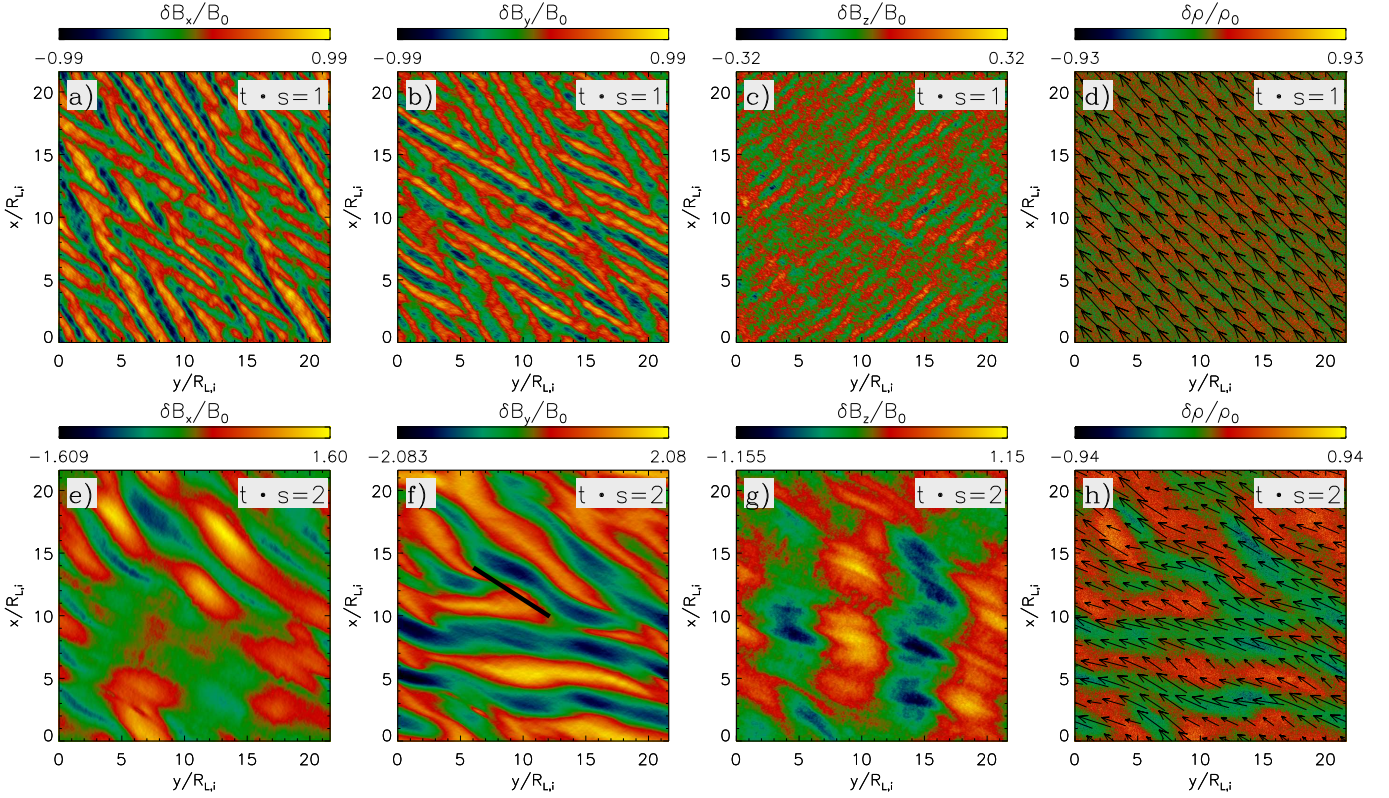}%{beta6.ps}
  \caption{The three components of $\delta \textbf{\textit{B}}$ and plasma density fluctuations $\delta \rho$ at two different times: $t\cdot s = 1$ (upper row) and $t\cdot s = 2$ (lower row), for run MW3 with $m_i/m_e = 128$. Fields and density are normalized by $B_0$ and the initial density $\rho_0$, respectively. Arrows in panels $d$ and $h$ show the mean magnetic field direction on the simulation $x-y$ plane. At both times, the magnetic fluctuations are dominated by oblique mirror modes, especially $\delta B_x$ and $\delta B_y$. Parallel propagating lower-amplitude IC modes are also apparent, particularly in $\delta B_z$, but are subdominant relative to the mirror modes.  Significantly shorter wavelength parallel propagating whistler modes are also present in $\delta B_z$; these have wavelengths shorter than that of the IC modes by a factor of $R_{L,i}/R_{L,e} \approx (m_i/m_e)^{1/2}=11$. The short black line in panel $f$ shows a small region where the whistler modes are apparent. The corresponding fluctuations are shown in Figure \ref{fig:fftwhistler}$b$.}
\label{fig:fldsmirrorwhistler}
\end{figure*}
\begin{figure}%[t!]
\centering
\includegraphics[width=9.2cm]{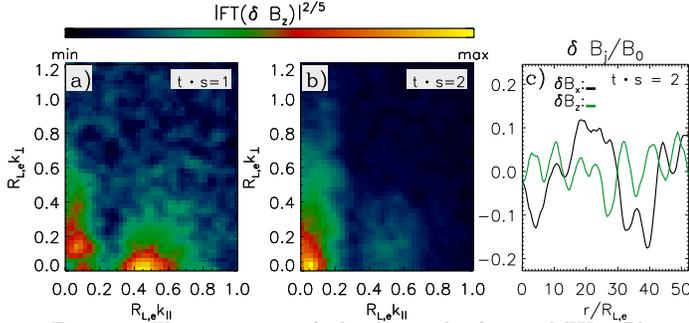}%{beta6.ps}
\caption{The signatures of whistler modes for run MW3. {\it Plot $a$:} the magnitude of the Fourier transform of $\delta B_z$ at $t\cdot s=1$ (raised to the $2/5$th power, $|\textrm{FT}(\delta B_z)|^{2/5}$, to provide better dynamical range), as a function of the wavenumbers parallel and perpendicular to $\langle \textbf{\textit{B}} \rangle$ ($k_{||}$ and $k_{\perp}$, respectively). The contribution from quasi-parallel whistler modes with wavevectors satisfying $kR_{L,e} \sim 0.5$, along with the subdominant contributions of the longer wavelength IC and mirror modes (quasi-parallel and oblique, respectively) are clearly seen. {\it Plot b:} same as in plot {\it a}, but at $t\cdot s=2$. In this case, most of the power is provided by the IC and mirror modes, with the whistler modes contributing subdominant power. {\it Plot c:} the $\delta B_x$ and $\delta B_z$ components of $\delta \textbf{\textit{B}}$ in a small one-dimensional region of the run at $t \cdot s =2$, revealing the presence of whistler modes with $kR_{L,e}\sim 0.5$. The location of the 1D region of this plot in the larger 2D computational plane is shown with the short black line in Figure \ref{fig:fldsmirrorwhistler}$f$.}
\label{fig:fftwhistler}
\end{figure}
\newline

Figure \ref{fig:cgl_1000and2500} shows the initial evolution (until $t \cdot s = 1$) of the electron pressures perpendicular (black-solid) and parallel (red-solid) to $\textbf{\textit{B}}$ for two runs, one with $\omega_{c,e}/s=1000$ and one with $\omega_{c,e}/s=2500$ (runs OW2 and OW1 in Table \ref{table:1D}, respectively). The black- and red-dotted lines show the expectation from the CGL or double adiabatic limit \citep{ChewEtAl56}, which is reasonably satisfied until $t \cdot s \sim 0.4$ in both simulations. After that, the growth of whistler modes provide enough pitch-angle scattering to break the adiabatic evolution of the electron pressure.\newline

Figure \ref{fig:benandanis_wces1000and5000} shows the time evolution of the magnetic fluctuations, the volume averaged pressure anisotropy, and the electron magnetic moment for the same runs until $t \cdot s =3$.  Panels $c$ and $d$ show the volume averaged pressure anisotropy $\langle \Delta p_e\rangle /\langle p_{||,e}\rangle $ for these two runs.  For comparison, in both cases we plot the electron pressure anisotropy that would produce a whistler instability growth rate, $\gamma_{w}$, equal to 5 times the shearing rate ($\gamma_{w}=5s=5\times10^{-3}\omega_{c,e}$ and $\gamma_w=5s=2\times 10^{-3}\omega_{c,e}$, respectively). These thresholds were obtained using the linear Vlasov solver developed by \cite{Verscharen2013} for mass ratio $m_i/m_e=1836.$\footnote{One subtlety is that the thresholds derived using the Vlasov solver of \cite{Verscharen2013} apply to the case of non-relativistic electrons. Since the electrons in our simulations are mildly relativistic ($kT_e=0.28 m_ec^2$), we used initial-value PIC simulations to find a calibration factor $f_c$ that scales the non-relativistic thresholds to the mildly relativistic regime (the threshold anisotropy for a given growth rate is larger by $f_c$ in the mildly relativistic regime). For the whistler growth rates and mass ratios used in the paper, we found that $f_c$ is a smooth function of $\beta_{||,e}$ only, with $f_c\approx 1.5$ and 2 for $\beta_{||,e}=5$ and 15, respectively. Thus the whistler thresholds in Figure \ref{fig:benandanis_wces1000and5000} and subsequent figures are the non-relativistic thresholds multiplied by $f_c$.} We see that in both cases there is a reasonably good agreement between the electron anisotropy obtained from the simulation (using $m_i/m_e=\infty$) and the theoretical whistler instability thresholds. This shows that the electron anisotropy is maintained at roughly the marginal stability level for the whistler modes with $\gamma_{w}=5s$.  Note that in a realistic astrophysical or heliospherical environment, $\omega_{c,e}/s$ is likely to be larger than in the simulations shown in Figure \ref{fig:benandanis_wces1000and5000}.  This would lead to somewhat lower saturated $\Delta p_e/p_{||,e}$ (see \S \ref{sec:whistlerandmirror}). \newline

The pressure anisotropy evolution in Figure \ref{fig:benandanis_wces1000and5000} is very similar for $\omega_{c,e}/s=1000$ and $\omega_{c,e}/s=2500$.  The primary difference between these two runs is in the amplitude of $\delta \textbf{\textit{B}}$. This is shown in panels $a$ and $b$, where we plot the magnetic energy density in the components of $\delta \textbf{\textit{B}}$ that point along and perpendicular to $\langle \textbf{\textit{B}}\rangle$ (normalized by the average magnetic energy density in the simulation, $\langle  B^2\rangle /8\pi$, which grows with time). We see first that $\delta \textbf{\textit{B}}$ is dominated by the component perpendicular to $\langle \textbf{\textit{B}}\rangle $, which is expected for the transverse nature of the whistler modes. Also, as the initial electron magnetization $\omega_{c,e}/s$ increases by a factor 2.5, the maximum value of $\delta \textbf{\textit{B}}^2$ decreases by a factor $\sim 1.5$. This behavior is consistent with the expectation that the whistler modes should produce an effective pitch angle scattering rate $\nu_{eff}$ proportional to $\omega_{c,e}^2 (\delta \textbf{\textit{B}}/B)^2$ \citep[see, e.g., ][]{Marsch2006}. Indeed, under the assumption that $\nu_{eff}$ should maintain the electron pressure anisotropy at the marginally stable level for whistler growth (which happens for $t\cdot s \gtrsim 0.7$ in Figures \ref{fig:benandanis_wces1000and5000}$c$ and \ref{fig:benandanis_wces1000and5000}$d$), the value of $\nu_{eff}$ can be estimated from the macroscopic properties of the flow
as follows:\newline

Let us consider the evolution of $p_{\perp,e}$ and $U_e=p_{\perp,e} + p_{||,e}/2$ in the case of an incompressible fluid, with homogeneous $p_{\perp,e}$ and $p_{||,e}$, and without heat flux along $\langle \textbf{\textit{B}}\rangle $, which is given by \citep{Kulsrud1983, SnyderEtAl1997, SharmaEtAl07}:
\begin{equation}
\frac{\partial p_{\perp,e}}{\partial t}=-sp_{\perp,e}B_xB_y/B^2 -\frac{1}{3}\nu_{eff} (p_{\perp,e}-p_{||,e})
\label{eq:1}
\end{equation}
and
\begin{equation}
\frac{\partial U_{e}}{\partial t}=-s\Delta p_{e}B_xB_y/B^2. 
\label{eq:2}
\end{equation} 
If $\Delta p_e/p_{||,e} \ll 1$ (which occurs in our case), $U_{e} \approx 3p_{\perp,e}/2$. Thus, assuming $B_x \sim |B_y| \sim B$ ($B_y < 0$ in our case), one gets that $\partial \textrm{ln}(U_e)/\partial t \sim \partial \textrm{ln}(p_{\perp,e})/\partial t \sim s \Delta p_e/p_{||,e} \ll s$. Equation \ref{eq:1} then implies that
\begin{equation}
\nu_{eff} \approx -3 s \frac{p_{||,e}}{\Delta p_e}\frac{B_xB_y}{B^2}.
\label{eq:nueff}
\end{equation}
Thus, comparing the $\nu_{eff}$ obtained in Equation \ref{eq:nueff} with the expected dependence of $\nu_{eff}$ on waves amplitude ($\nu_{eff} \propto \omega_{c,e}^2 (\delta B^2/B^2)$) one obtains:
\begin{equation}
\frac{\delta B^2}{B^2} \propto \frac{p_{||,e}}{\Delta p_e} \frac{|B_xB_y|}{B^2}\frac{s}{\omega_{c,e}^2}.
\label{eq:deltaB2}
\end{equation}
Given that runs OW1 and OW2 have the same initial $\omega_{c,e}$, the decrease in $s$ by a factor 2.5, along with the observed decrease in $\Delta p/p_{||,e}$ by a factor $\sim 1.5$ (see Figures \ref{fig:benandanis_wces1000and5000}$c$ and \ref{fig:benandanis_wces1000and5000}$d$), is fairly consistent with the decrease in $\delta B^2/B^2$ by a factor $\sim 1.5$. Also, the roughly constant behavior of $\delta B^2/B^2$ in the range $t\cdot s \sim 0.5-1.5$ (see Figures \ref{fig:benandanis_wces1000and5000}$a$ and \ref{fig:benandanis_wces1000and5000}$b$) is consistent with the fact that in the early stage  of the saturated regime both $p_{||,e}/\Delta p_e$ and $\omega_{c,e}^{-2}$ decrease slowly with time, which is nearly compensated by the initial growth of $|B_xB_y|/B^2$ ($\propto t$). For $t\cdot s \gg 1$, the expectation is $|B_xB_y|/B^2 \propto 1/t$ and $\omega_{c,e}^{-2} \propto 1/t^2$, consistent with the rapid decrease of $\delta B^2/B^2$ at the end of the simulations. This behavior implies the absence of a long-term secular growth of $\delta B^2/B_0^2$, as can be seen from the red-dotted lines in Figures \ref{fig:benandanis_wces1000and5000}$a$ and \ref{fig:benandanis_wces1000and5000}$b$.\newline
%giving a nearly constant behavior of  and then gradually starts decreasing, with $|B_xB_y|/B^2 \propto 1/t$ for $t \cdot s \gg 1$. Thus 
%For $\nu_{eff} \propto s$, as $\omega_{c,e}$ increases over time (due to $B$ amplification), whistler modes with lower amplitude are able to scatter the electrons; this explains why $(\delta \vec{B}/B)^2$ decreases with time in both the $\omega_{c,e}/s=1000$ and $\omega_{c,e}/s=2500$ cases (see Figures \ref{fig:benandanis_wces1000and5000}$a$ and \ref{fig:benandanis_wces1000and5000}$b$).\footnote{We show below that the scattering rate $\nu_{eff}$ necessary to maintain marginal stability is $\sim s \langle  p_{||,e}\rangle /\langle \Delta p_e\rangle $. Thus, if $\langle \Delta p_e\rangle /\langle p_{||,e}\rangle $ grows with time (as occurs here; see Figure \ref{fig:benandanis_wces1000and5000}), $\nu_{eff}$ should decrease, reinforcing the need for $\delta B^2/B^2$ to decrease as time goes on.}\newline

Finally, we define two magnetic moments to aid in interpreting the numerical results:
\begin{equation}
\langle \mu_j\rangle  \equiv \Big\langle \frac{p_{\perp,j}}{B}\Big\rangle  \ \ \ \ {\rm and} \ \ \ \  \mu_{j,eff} \equiv \frac{\langle p_{\perp,j}\rangle }{\langle B\rangle }
\label{eq:mu}
\end{equation}
$\langle \mu_j\rangle $ is the true volume averaged magnetic moment.  In \citet{RiquelmeEtAl2015} we showed that, in the case of the ions, $\mu_{i,eff} \ne \langle \mu_i \rangle $, which is produced when there is a spatial correlation between $p_{\perp,i}$ and $B$, as in the case of large amplitude mirrors.  Figures \ref{fig:benandanis_wces1000and5000}$e$ and \ref{fig:benandanis_wces1000and5000}$f$ compare these two definitions of the electron magnetic moment for the same runs OW2 and OW1. We see that for the two simulations, $\langle \mu_e\rangle $ decreases on the same time scale ($\sim s^{-1}$).  The fact that $\langle \mu_e\rangle $ and $\mu_{e,eff}$ are essentially indistinguishable in Figure \ref{fig:benandanis_wces1000and5000} means that $p_{\perp,e}$ does not fluctuate significantly in space.  This is consistent with the relatively low amplitude fluctuations in $\delta B$ associated with the whistler instability.  We will see below that this is no longer the case when mirror fluctuations are present.\newline

\subsection{Simulations With Whistler and Mirror Modes}
\label{sec:whistlerandmirror}

\noindent In order to study the interplay between the electron-scale whistler instability and the ion-scale mirror instability, we now study a series of simulations with finite ion to electron mass ratios $m_i/m_e$.  Ideally we would utilize $m_i/m_e \simeq 1836$ but this is infeasible given the need for both 2D and large ion and electron magnetization.  Instead, we have tried to ensure that the simulations are in the regime where there is reasonable scale separation between ions and electrons.  This is achieved for $m_i/m_e = 128$ but even at somewhat smaller mass ratios we find reasonably similar results. \newline

As an example, Figure \ref{fig:fldsmirrorwhistler} shows the components of $\delta \textbf{\textit{B}}$  for run MW3 of Table \ref{table:1D} ($m_i/m_e=128$ and $\omega_{c,e}/s=5000$).  The upper and lower rows correspond to $t\cdot s=1$ and $t\cdot s=2$, respectively. At $t\cdot s=1$ the oblique mirror modes are visible in $\delta B_x$ and $\delta B_y$, while the whistler modes are most clearly seen in Figure \ref{fig:fldsmirrorwhistler}c, which shows $\delta B_z$. At $t\cdot s=2$ we see a well developed highly nonlinear stage of all the instabilities. Whereas $\delta \textbf{\textit{B}}$ is dominated by the mirror modes (with wavenumber $k$ such that $kR_{L,i}\sim 1$, where $R_{L,i}$ is the ion Larmor radius), the $\delta B_z$ component also shows the (subdominant) presence of the IC modes. This is consistent with our previous results \citep{RiquelmeEtAl2015}: although subdominant, the IC instability persists for $\beta_i \sim 10$ (though it is not present at higher $\beta_i$). Figure \ref{fig:fldsmirrorwhistler}$h$ also shows significant plasma density fluctuations, which correlate well with the mirror modes. At $t\cdot s=2$, the three components of $\delta \textbf{\textit{B}}$ also show the presence of parallel whistler modes on scales comparable to $R_{L,e} \approx R_{L,i}/11$ (consistent with $\sqrt{m_i/m_e}\approx11$).  

The presence of the different ion- and electron-scale modes can also be seen from Figure \ref{fig:fftwhistler}. Figures \ref{fig:fftwhistler}$a$ and \ref{fig:fftwhistler}$b$ show the Fourier transform of $\delta B_z$ at $t\cdot s=1$ and $t\cdot s=2$, respectively, as a function of the wavenumbers parallel and perpendicular to $\langle \textbf{\textit{B}} \rangle$. At $t\cdot s=1$, the quasi-parallel whistler modes with $kR_{L,e} \sim 0.5$ contribute most of the power, with the contribution of smaller wavenumber, quasi-parallel and oblique modes (IC and mirror modes, respectively) being subdominant. At $t\cdot s=2$, the whistler, IC, and mirror modes continue to contribute $B_z$ fluctuations in similar regions of the $k_{||}-k_{\perp}$ space, but with the whistler modes having significantly less power compared to the IC and mirror modes. Figure \ref{fig:fftwhistler}$c$ shows $B_x$ and $B_z$ fluctuations in a small one-dimensional region (marked by a small black line in Figure \ref{fig:fldsmirrorwhistler}$f$) at $t \cdot s =2$. The presence of whistler modes with $kR_{L,e}\sim 0.5$ appears clearly as low amplitude fluctuations (relative to the mirror modes).\newline
\begin{figure}[t!]
\centering
\includegraphics[width=9cm]{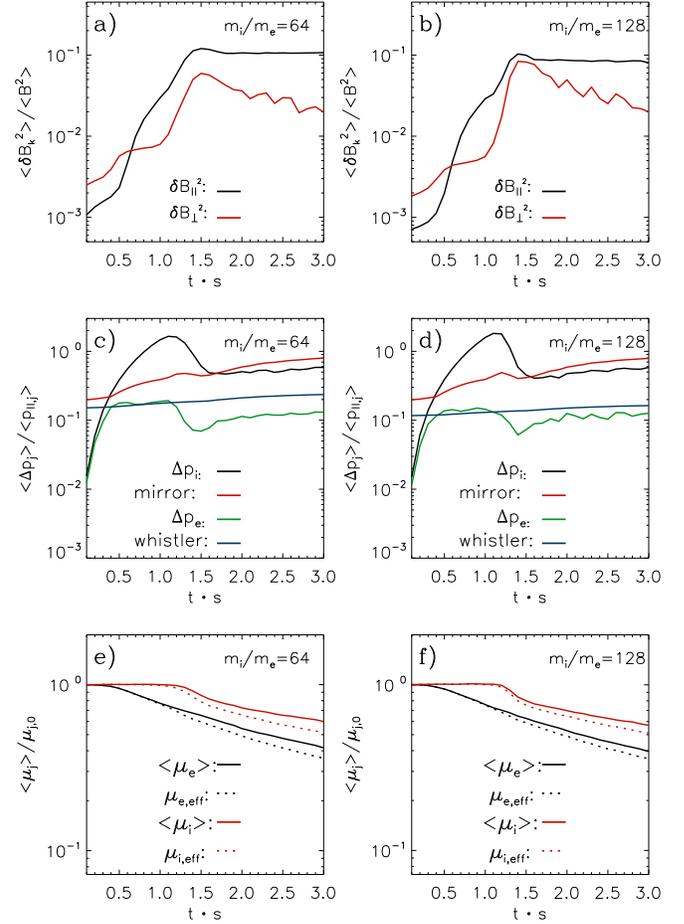}%{beta6.ps}
\caption{Time evolution of volume-averaged properties for simulations with $m_i/m_e=64$ (left column; run MW2) and $m_i/m_e=128$ (right column; run MW3).    The upper row shows the evolution of the magnetic energy parallel (black) and perpendicular (red) to $\langle \textbf{\textit{B}}\rangle $, normalized by $B^2/8\pi$. The middle row shows the ion (black) and electron (green) pressure anisotropies, $\Delta p_j/p_{||,j}$. Panels $c$ and $d$ also contain the anisotropy thresholds for mirror (red) and whistler (blue) modes growing at growth rates of $\sim s$ (using $m_i/m_e=64$ and 128, respectively).   The electron pressure anisotropy saturates at a value $\sim 2$ times lower than the value expected if the isotropization were dominated by whistler modes only; this is due to bunching by the large-amplitude mirrors generated by the ions.   The lower row shows the ion (red) and electron (black) magnetic moments, defined as in equation \ref{eq:mu}, and normalized by the initial value of $\mu_j$.}
\label{fig:benandanis_wces2500_mime25and64}
\end{figure}

\noindent Figure \ref{fig:benandanis_wces2500_mime25and64} compares the evolution of the energy in $\delta \textbf{\textit{B}}$, the ion and electron anisotropies, and $\mu_i$ and $\mu_e$ for simulations with $m_i/m_e=64$ and 128 (runs MW2 and MW3 in Table \ref{table:1D}), and demonstrates that the physics of electron isotropization is fairly well converged for these two mass ratios. Although these two simulations differ in $m_i/m_e$, the ions are under the same conditions ($\omega_{c,i}/s=40$ and $\beta_{i}=20$ in the two cases).  The electrons are also under similar conditions (the same $\beta_{e}=20$, $k_BT_e=0.28 m_e c^2$), but their magnetizations differ by a factor 2 (see Table \ref{table:1D}), which is required by the factor 2 difference in $m_i/m_e$. Figures \ref{fig:benandanis_wces2500_mime25and64}a and \ref{fig:benandanis_wces2500_mime25and64}b show the magnitude of the volume-averaged magnetic energy in fluctuations parallel and perpendicular to the volume-averaged (shearing) magnetic field, $\langle \textbf{\textit{B}}\rangle (t)$, normalized by $\langle B^2\rangle $. We see that, for both mass ratios, the amplitude of the mirror modes is about the same at saturation, confirming the results of \cite{Kunz2014} and \cite{RiquelmeEtAl2015} that $\delta B/B \sim 0.3$ in the saturated mirror state. There is also a subdominant perpendicular field component, $\delta B_{\perp}^2$, which is at first dominated by the whistler modes (most clearly seen by the exponential growth at $t\cdot s \approx 0.5$ in Figures \ref{fig:benandanis_wces2500_mime25and64}a and \ref{fig:benandanis_wces2500_mime25and64}b), and then by the IC modes at later times.\newline

Figures \ref{fig:benandanis_wces2500_mime25and64}$c$ and \ref{fig:benandanis_wces2500_mime25and64}$d$ show the volume-averaged electron and ion pressure anisotropies as a function of time (green and black lines, respectively).  The anisotropy evolution for the two species is essentially the same for the two mass ratios, although there are small quantitative differences. Figures \ref{fig:benandanis_wces2500_mime25and64}$c$ and \ref{fig:benandanis_wces2500_mime25and64}$d$ also show the anisotropy threshold for mirror (red line) and whistler modes (blue line).  The linear theory mirror instability threshold for growth rate equal to $s$ reasonably describes the saturation of the ion pressure anisotropy (aside from the overshoot at $t \cdot s \sim 1$ that is unavoidable at the modest ratios of the ion-cyclotron frequency to the shear rate used here). However, the electron pressure anisotropy is a factor $\sim 1.5-2$ smaller than the linear theory threshold for whistler modes.\footnote{In the case of whistler modes, the theoretical thresholds correspond to growth rates of $5s$, since these are the rates that fit fairly well the electron anisotropies in the case of ``infinite mass" ions (see Figures \ref{fig:benandanis_wces1000and5000}$c$ and \ref{fig:benandanis_wces1000and5000}$d$).} This suggests that the factor $\sim 1.5-2$ decrease in $\Delta p_{e}/p_{||,e}$ is caused by the presence of the mirror modes in the simulations with finite mass ratios (runs MW2 and MW3).\newline
\begin{figure}[t!]
  \centering
  \includegraphics[width=8cm]{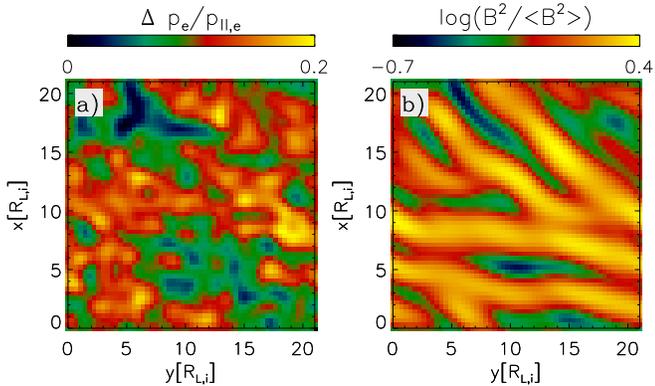}%{beta6.ps}
  \caption{Comparison of the spatial distributions of $\Delta p_e/p_{||,e}$ (panel $a$) and $B^2$ (panel $b$) for run MW3 at $t\cdot s = 2$ (same run and time shown in the lower row of Figure \ref{fig:fldsmirrorwhistler}).}
\label{fig:aniso2d}
\end{figure}

\noindent This reduction in $\Delta p_e/p_{||,e}$ can be understood in terms of the effect of the nonlinear mirror modes on the electrons.  The mirrors tend to bunch electrons (and ions) into low magnetic energy regions, which contributes to reducing the volume-averaged anisotropy. This can also be seen from panels \ref{fig:benandanis_wces2500_mime25and64}e and \ref{fig:benandanis_wces2500_mime25and64}f, where we compare $\langle \mu_j\rangle  =\langle p_{\perp,j}/B\rangle $ and $\mu_{j,eff} =\langle p_{\perp,j}\rangle /\langle B\rangle $ for both ions and electrons.  We see that $\mu_{j,eff}$ tends to be noticeably smaller than $\langle \mu_j\rangle $ (by $\sim 20\%$) at $t\cdot s \gtrsim 1$, implying that the mirror modes partially reduce $p_{\perp,j}$ in a way that conserves $\mu_j$ (bunching them into mirrors).\newline
 
\noindent This effect can also be seen from Figure \ref{fig:aniso2d}, which compares the spatial distributions of $\Delta p_e/p_{||,e}$ (Figure \ref{fig:aniso2d}$a$) and $B^2$ (Figure \ref{fig:aniso2d}$b$) for run MW3 at $t\cdot s = 2$ (the same run and time shown in the lower row of Figure \ref{fig:fldsmirrorwhistler}). The presence of significant $\Delta p_e/p_{||,e}$ fluctuations on scales comparable to the typical scales of the (mirror-dominated) $B^2$ variations underscores the importance of the mirror modes in regulating $\Delta p_e/p_{||,e}$. The effect of mirror modes on $<\Delta p_e>/<p_{||,e}>$ in the $m_i/m_e=128$ case, however, is smaller than in the $m_i/m_e=64$ case, suggesting that the mirrors would have less of an effect on the electron anisotropy (relative to the whistler modes) at even larger $m_i/m_e$. We thus consider the factor $\sim 1.5-2$ an upper limit for the effect of mirror modes on the electron anisotropy.\newline

One noticeable difference between the ion and electron response to the growing magnetic field is that $\mu_i$ is conserved to reasonably high accuracy for $t \cdot s \lesssim 1$, while $\mu_e$ is not (Figures \ref{fig:benandanis_wces2500_mime25and64}$e$ and \ref{fig:benandanis_wces2500_mime25and64}$f$).   This is because the mirror instability has a secular phase that conserves $\mu_i$ \citep{SchekochihinEtAl05,Kunz2014}.   The adiabatic invariance of $\mu_i$ is only broken when the mirrors reach $\delta B \sim B$, which happens at $t \cdot s \sim 1$.   By contrast, the electron magnetic moment decreases at much earlier times.  This is due to the electron whistler instability which does not have a secular phase and which can pitch angle scatter the electrons at low amplitudes, and hence at $t \cdot s \lesssim 1$.

%Indeed, if the electron pressure isotropization were completely dominated by this effect, $\mu_{e,eff}$ would still decrease on a timescale of $s^{-1}$ while $<\mu_e>$ would be constant in time.  
%This contribution of the mirror modes to the isotropization of partcles' pressure is also observed in the case of the ions (see solid-red and dotted-red lines in panels \ref{fig:benandanis_wces2500_mime25and64}e and \ref{fig:benandanis_wces2500_mime25and64}f), which reproduces the result of \cite{RiquelmeEtAl2015}. \newline
 \begin{figure}[t!]
  \centering
  \includegraphics[width=9.8cm]{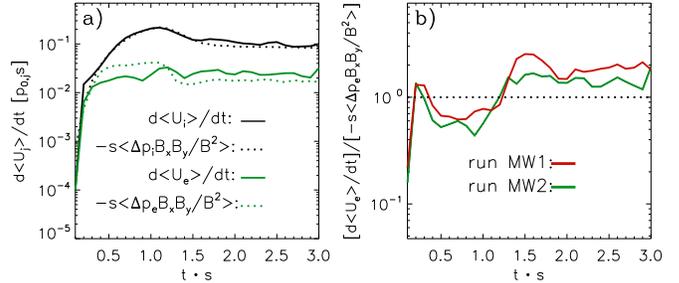}%{beta6.ps}
  \caption{{\it Panel $a$:} the ion and electron heating rates for run MW2, calculated directly via $d\langle U_i\rangle /dt$ and $d\langle U_e\rangle /dt$ (solid black and green, respectively), where $U_j$ is the internal energy per unit volume of species $j$ .  For comparison, we also show that the numerically calculated heating rates are well explained by the theoretically predicted ion (dotted-black) and electron (dotted-green) energy gain via `anisotropic viscosity' tapping into the background velocity shear (see eq. \ref{eq:2}). {\it Panel $b$:} the ratio between the electron heating, $d\langle U_e\rangle /dt$, and the expectation from viscous heating for runs MW1 (red) and MW2 (green), which only differ in their $N_{ppc}$ ($=20$ and $60$, respectively).}
\label{fig:energygainwhismirro}
\end{figure}
\newpage

\subsection{Viscous Heating}
\label{sec:viscosity}

\noindent Figure \ref{fig:benandanis_wces2500_mime25and64} demonstrates the existence of a quasi-steady state pressure anisotropy at a level set by the threshold of the mirror/whistler instabilities.  This in turn corresponds to an effective viscosity for both ions and electrons: in the present context, $P_{xy} \propto (p_\perp - p_\parallel) B_x B_y/B^2$ since the particles are roughly gyrotropic in velocity-space, where $P_{xy}$ is the $x-y$ component of the pressure tensor.  In our simulations, this anisotropic pressure can tap into the velocity shear in the plasma, converting shear energy into random thermal energy.  To quantify the importance of this heating mechanism in our simulations, Figure \ref{fig:energygainwhismirro}$a$ shows the volume-averaged ion (solid-black) and electron (solid-geen) heating rates for run MW2: $d \langle U_j\rangle /dt$, where $U_j$ is the internal energy per unit volume of species $j$. For comparison, we also plot the expected ion (dotted-black) and electron (dotted-green) heating rates due to the work done by anisotropic viscosity: $-s \langle \Delta p_j B_x B_y/B^2\rangle$, which is obtained from equation \ref{eq:2}.\footnote{We compared the heating predicted by two different volume averages: $-s\langle \Delta p B_xB_y/B^2\rangle $ and $-s\langle \Delta p\rangle \langle B_x\rangle \langle B_y\rangle /\langle B^2\rangle $.  The results were nearly indistinguishable at all times.  This implies that the correlations in the fluctuating fields do not significantly change the heating rate in these calculations, even in the presence of large amplitude mirrors.} \newline

\noindent For both for ions and electrons there is good agreement between the particle heating in the simulation and the contribution from the anisotropic stress. In the case of the electrons the measured heating is moderately larger (by a factor $\sim 1.5$) than the viscous heating expectation. This can be seen more clearly in Figure \ref{fig:energygainwhismirro}$b$, which shows the ratio between the measured electron heating and the expected contribution from anisotropic viscosity for run MW2 (green line). For comparison, we also show the case of run MW1 (red line), which uses N$_{\textrm{ppc}}$=20 (instead of N$_{\textrm{ppc}}$=60, as in run MW2; all the other parameters are the same). The fact that run MW1 shows an extra factor $\sim 1.5$ increase in the electron heating suggests that a significant contribution to the additional heating comes from the numerical noise due to the limited values of N$_{\textrm{ppc}}$ feasible in our simulations. \newline 

\noindent It has also been argued that the energy transfer from ion-scale turbulent fluctuations to the electrons could be a possible source of electron heating (\citealt{SironiEtAl2015a}, though in a regime where the ion-cyclotron instability dominates over the mirror instability; see \S \ref{sec:conclu}). Another possible factor at play is that the electrons' energy could be somewhat reduced by the growth of the waves' energy itself. This effect appears to be significant in the early part of the simulations, when the energy content in the mirror/IC/whistler fluctuations is rapidly growing. Indeed, Figure \ref{fig:energygainwhismirro} shows that for $t\cdot s \lesssim 1.2$ the heating rate of the electrons is somewhat smaller than the viscous heating prediction. However, despite these considerations, the electron heating obtained for run MW2 shows that the anisotropic viscosity constitutes the dominant mechanism for electron heating in our simulations.

\section{Electron Mean Free Path}
\label{sec:conduct}

\noindent The pitch-angle scattering created by velocity-space instabilities provides an upper limit to the particles' mean free path in a low collisionality plasma.  This in turn determines the magnitude of the effective viscosity and thermal conductivity in the plasma.  In this section we quantify this directly by computing the ion and electron mean free paths ($\lambda_i$ and $\lambda_{e}$) along the mean magnetic field, $\langle \textbf{\textit{B}}\rangle $, during the nonlinear stage of the whistler and mirror instabilities. In order to do so, in each simulation we compute the distance $D_j(t)$ traveled along $\langle \textbf{\textit{B}}\rangle $ for $2\times 10^4$ ions and electrons.\footnote{$D_j(t)\equiv \int_{0}^t \textbf{\textit{v}}_j\cdot \textbf{\textit{B}}/B dt$, where $\textbf{\textit{v}}_j$ is the particle's velocity.} If the particle trajectories are random walks, then $\langle D_j^2\rangle = tv_{th,j}\langle \lambda_j \rangle$ (where $\langle \lambda_j \rangle$ represents the average mean free path over species $j$, and $v_{th,j}=(k_BT_j/m_j)^{1/2}$ is the thermal speed).  Calculating $d\langle D_j^2\rangle/dt$ then gives an estimate of the average mean free path $\langle \lambda_j \rangle$ of species $j$. As in \S \ref{sec:interplay}, we first describe our calculation of $\langle \lambda_e \rangle$ for simulations with infinite mass ions. This way we will clearly separate the effect of mirror and whistler modes on $\langle \lambda_{e} \rangle$.\newline 
\begin{figure}[t!]  \centering \includegraphics[width=7cm]{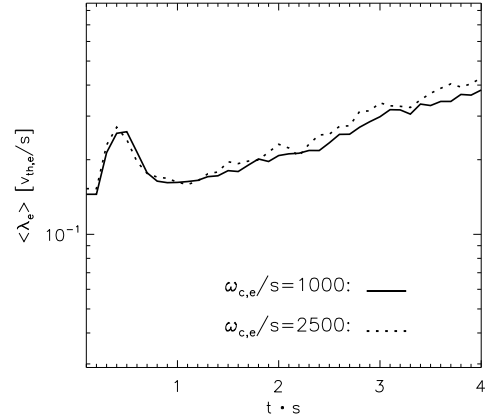} \caption{The average electron mean free path, $\langle \lambda_e \rangle$, normalized by $v_{th,e}/s$ and calculated via the time derivative of the mean squared distance traveled by electrons along the mean magnetic field $\langle \textbf{\textit{B}}\rangle $, for runs with infinite ion mass and with $\omega_{c,e}/s=1000$ (solid; run OW2) and $\omega_{c,e}/s=2500$ (dotted; run OW1). There is no significant dependence of the estimated mean free path on the magnetization $\omega_{c,e}/s$.    Aside from an early free streaming phase, the mean free path is well estimated via $\langle \lambda_e \rangle \approx 0.3 (\langle \Delta p_e \rangle/\langle p_{||,e}\rangle)(B^2/|B_xB_y|) v_{th,e}/s$ (see eqs \ref{eq:1} \& \ref{eq:2} and associated discussion). The late-time increase in the electron mean free path is consistent with the increase in $\langle \Delta p_e \rangle /\langle p_{||,e}\rangle$ and $B^2/|B_xB_y|$.}  \label{fig:mfpmime2000} \end{figure} 
\begin{figure}[t!]  
\centering 
\includegraphics[width=7cm]{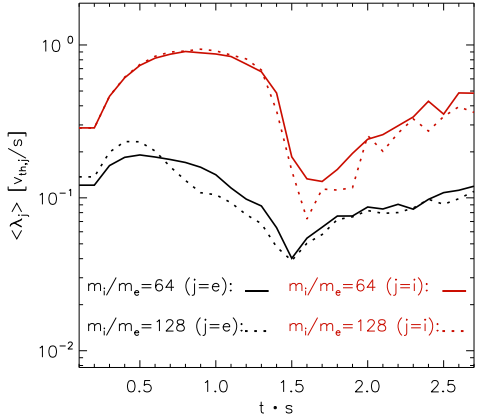} 
\caption{Electron (black) and ion (red) mean free paths (normalized by $v_{th,j}/s$), calculated via the time derivative of the mean squared distance traveled by particles along $\langle \textbf{\textit{B}}\rangle $.   We show results for runs with $m_i/m_e=64$ (solid lines; run MW2) and $m_i/m_e=128$ (dotted lines; run MW3).  At early times the particles undergo a period of free-streaming in which the inferred mean free path increases.   After the velocity-space instabilities saturate, however, pitch angle scattering ensues leading to a rough saturation of the mean free path. The simulations with different mass ratios give similar results, with $\langle \lambda_j \rangle \approx 0.3 (\langle \Delta p_j \rangle /\langle p_{||,j}\rangle)(B^2/|B_xB_y|) v_{th}/s$ in both cases (see Equations \ref{eq:1} \& \ref{eq:2} and associated discussion).}  
\label{fig:mfpwces2500} 
\end{figure}

\noindent Figure \ref{fig:mfpmime2000} shows $\langle \lambda_e \rangle \equiv d\langle D_e^2\rangle /dt/v_{th,e}$ (normalized by $v_{th,e}/s$) for simulations with infinite mass ions and for electron magnetizations, $\omega_{c,e}/s=1000$ and 2500 (simulations OW2 and OW1, respectively).   The evolution of $\langle \lambda_{e}\rangle$ for the two electron magnetizations is very similar.
%Since the expected behavior for electrons that follow random walk trajectories is $<D^2>_e\approx \lambda_{e}v_{th,e}t$, then $d<D^2>_e/dt/(v_{th,e}^2/s)$ is effectively measuring $\lambda_{e}v_{th,e}$
%Figure \ref{fig:mfpmime2000} shows that the evolution of $\lambda_{e}s/v_{th,e}$ for the $\omega_{c,e}/s=1000$ and 5000 cases is essentially the same. 
At the beginning there is a small period of time when $\langle \lambda_{e} \rangle$ increases rapidly $\sim t$. This is consistent with an initial ``free streaming" of the electrons (in which $d\langle D_e^2\rangle /dt \propto t$), followed by a sudden decrease in the mean free path due to the strong scattering at the end of the exponential whistler growth phase (where a transient anisotropy ``overshoot" occurs, leading to an overshoot in the mean free path; see Figures \ref{fig:benandanis_wces1000and5000}c and \ref{fig:benandanis_wces1000and5000}d, corresponding to the same runs OW2 and OW1).  By $t \cdot s \sim 0.5$, the whistler modes have reached the fully saturated regime, and $\langle \lambda_{e} \rangle \sim 0.15 v_{th,e}/s$. After that, $\langle \lambda_{e} \rangle$ grows with time, increasing by a factor of $\sim 2$ by the end of the simulation.
%As we now describe, this can also be interpreted as the effective collision rate, $\nu_{eff}$ (due to whistler-driven pitch-angle scattering) being $\nu_{eff} \sim 20s$ at the beginning of the simulations, and decreasing by a factor $\sim 3$ by $t \cdot s=4$. 
\newline

\noindent The numerically determined evolution of $\langle \lambda_e \rangle$ can be understood by considering the expression for $\nu_{eff}$ given by Equation \ref{eq:nueff}, which is equivalent to $\langle \lambda_e \rangle \approx 0.3 (\langle \Delta p_e \rangle /\langle p_{||,e} \rangle) (B^2/|B_xB_y|) v_{th,e}/s$. At $t \cdot s =1$, the simulations with fixed ions have $\langle \Delta p_e \rangle/ \langle p_{||,e} \rangle \simeq 0.15$ (Figures \ref{fig:benandanis_wces1000and5000}$c$ and \ref{fig:benandanis_wces1000and5000}$d$) and $B^2/|B_xB_y| \simeq 2$, which corresponds to $\langle \lambda_e \rangle \approx 0.15$ $v_{th,e}/s$.  This is in reasonable agreement with the numerically determined values in Figure \ref{fig:mfpmime2000}.  The factor $\sim 2$ increase in the electron mean free path from $t \cdot s =1$ to $t \cdot s =3$ in Figure \ref{fig:mfpmime2000} is consistent with the factor $\sim 1.5$ increase in both $\Delta p_e/p_{||,e}$ (due to the decreasing $\beta_e$, which increases the threshold pressure anisotropy for the whistler instability) and in $B^2/|B_xB_y|$  (from $\sim 2$ to $\sim 3$).\newline

\noindent Figure \ref{fig:mfpwces2500} shows our calculations of the electron and ion mean free paths for simulations with $m_i/m_e=64$ and $128$ (runs MW2 and MW3, respectively).  Relatively independent of the mass ratio, the net effect of the mirror modes is to reduce the electron mean free path by a factor $\sim 2$ relative to the whistler-only results in Figure \ref{fig:mfpmime2000}. Since the mirror modes also reduce $\langle \Delta p_e \rangle/ \langle p_{||,e} \rangle$ by a factor of $\sim 1.5-2$, the result $\langle \lambda_e \rangle \approx 0.3 (\langle \Delta p_e \rangle/\langle p_{||,e} \rangle) (B^2/|B_xB_y|) v_{th,e}/s$ derived above continues to describe the behavior of $\langle \lambda_e \rangle$ when mirror modes are present. Figure \ref{fig:mfpwces2500} also shows the inferred average ion mean free path, $\langle \lambda_{i} \rangle s/v_{th,i}$, which is a factor of $\sim 3$ larger than that of the electrons.  This is consistent with both species satisfying $\langle \lambda_j \rangle \approx 0.3 (\langle \Delta p_j \rangle/\langle p_{||,j}\rangle) (B^2/|B_xB_y|) v_{th,j}/s$ given that $\langle \Delta p_i\rangle/\langle p_{||,i}\rangle$ is a factor $\sim 3$ larger than $\langle \Delta p_e\rangle/\langle p_{||,e}\rangle$ in our simulations (see Figures \ref{fig:benandanis_wces2500_mime25and64}$c$ and \ref{fig:benandanis_wces2500_mime25and64}$d$).\newline
\begin{figure}[t!]  
\centering 
\includegraphics[width=6.5cm]{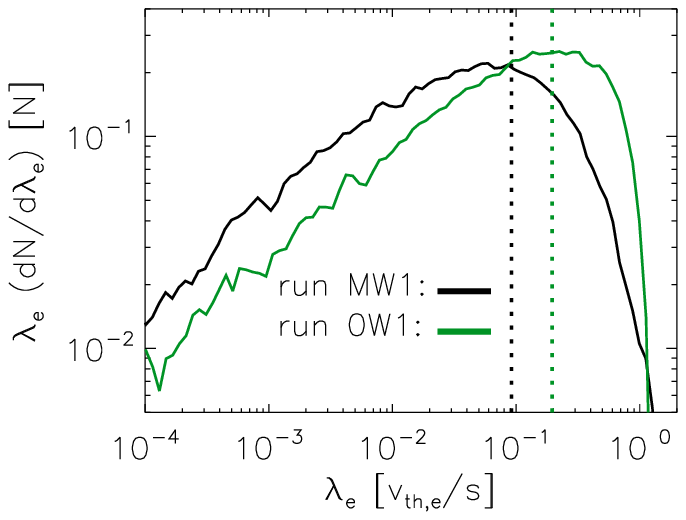} 
\caption{The distribution of electron mean free paths, $\lambda_e (dN/d\lambda_e)$, for electrons in runs OW1 (only whistlers; green line) and MW1 (whistlers and mirrors; black line), normalized by the total number of particles ($N$; $(dN/d\lambda_e)d\lambda_e$ is the number of particles with mean free path between $\lambda_e$ and $\lambda_e+d\lambda_e$). $\lambda_e$ for each particle is calculated measuring the distance $d$ traveled by each electron along \textbf{\textit{B}} during an interval 
$\Delta t=s^{-1}$, from $t\cdot s=1.5$ to 2.5, and assuming $d^2=\lambda_e v_{th,e} \Delta t$. The vertical-dotted green and black lines mark the average $\langle \lambda_e \rangle$ for the OW1 and MW1 runs, respectively.}  
\label{fig:2mfphisto} 
\end{figure}

\noindent Figure \ref{fig:2mfphisto} compares the probability distributions of mean free paths, $\lambda_e$, for electrons in runs OW1 (only whistlers; green line) and MW1 (whistlers and mirrors; black line). This is done by measuring the distance $d$ traveled by each electron along \textbf{\textit{B}} during an interval $\Delta t=s^{-1}$, from $t\cdot s=1.5$ to 2.5. This allows to estimate $\lambda_e$ for each individual electron by assuming $d^2=\lambda_e v_{th,e} \Delta t$.\footnote{We have chosen the interval $t\cdot s=1.5$ to 2.5 so that: $i)$ the mirror modes in run MW1 are in the fully saturated regime, and $ii)$ $\Delta t=s^{-1}$ is much larger than the average pitch-angle scattering time, $\sim \nu_{eff}^{-1}$, of electrons (necessary to assume diffusion). According to Equation \ref{eq:nueff},  $\nu_{eff}^{-1} \sim 0.1s^{-1}$, so we can safely assume $d^2=\lambda_e v_{th,e} \Delta t$.} The vertical-dotted green and black lines mark the average $\langle \lambda_e \rangle$ for the OW1 and MW1 runs, respectively, and reproduce the factor $\sim 2$ difference between the cases with and without mirrors (shown in Figures \ref{fig:mfpwces2500} and \ref{fig:mfpmime2000}, respectively). The effect of the mirror modes is to shift the $\lambda_e$ distribution to lower values of $\lambda_e$ (by a factor $\sim 2-3$, as seen in Figure \ref{fig:2mfphisto}). This can be understood as the electrons experiencing both pitch-angle scattering by whistler waves and trapping by large amplitude mirror modes. Pitch-angle scattering tends to untrap the trapped electrons by taking them into the loss cone of the mirror modes on time scales comparable to the mean pitch-angle scattering time, $\sim \nu_{eff}^{-1}$. Pitch-angle scattering can also trap the untrapped electrons on similar time scales \citep[see, e.g., ][]{KomarovEtAl16}. Thus, the distance $d$ traveled by an electron during a time $\Delta t$ (and, therefore, the estimated value of $\lambda_e$) should be scaled down by a factor that reflects the fraction of the time that the electrons are untrapped and free to move diffusively.

\section{Discussion and Implications}
\label{sec:conclu}

\noindent We have used particle-in-cell (PIC) plasma simulations to study the nonlinear evolution of ion and electron velocity-space instabilities in collisionless plasmas. We have focused on instabilities driven by pressure anisotropy with $p_{\perp,j} > p_{\parallel,j}$.  The motivation for doing so is in part that this sign of pressure anisotropy describes the typical conditions found in simulations of low-collisionality accretion flows onto black holes \citep{SharmaEtAl06,RiquelmeEtAl2012, FoucartEtAl16}.  In our calculations, an imposed shear velocity in the plasma amplifies a background magnetic field on a timescale long compared to the cyclotron motion of particles.  This drives $p_{\perp,j} > p_{\parallel,j}$ by the adiabatic invariance of the magnetic moment.  The pressure anisotropy in turn drives velocity-space instabilities.  The nonlinear, saturated state then depends on how the velocity-space instabilities inhibit the growth of pressure anisotropy.  \newline

\noindent In order to achieve reasonable scale separation between ions and electrons, we have focused on moderately large values for the mass ratio $m_i/m_e$, and found that for $m_i/m_e=64$ and $m_i/m_e=128$ our results are fairly independent of the mass ratio.  Our calculations have focused on the parameter regime $\beta_i \approx \beta_e=1-20$, which is relevant for a wide variety of heliospheric and astrophysical plasmas.  In particular, our simulations begin with $\beta_i = \beta_e=20$ but $\beta_j$ decreases as the background magnetic field is amplified in time.  In this regime the mirror instability is the dominant ion-scale instability (although with a subdominant contribution from the IC instability).  As in previous work \citep{Kunz2014,RiquelmeEtAl2015}, the mirror instability grows to large amplitudes $\delta B \sim 0.3 B$, even when the background magnetic field is amplified on a timescale long compared to the ion/electron cyclotron periods.  In addition to the mirror instability, the anisotropic electrons excite the whistler instability on scales of order the electron Larmor radius, much smaller than the scale of the mirror modes (see Figures \ref{fig:fldsmirrorwhistler} and \ref{fig:fftwhistler}).\newline
% dominates the electron isotropization, with a moderate contribution from the mirror modes, 

\noindent In the nonlinear saturated state, the ion and electron pressure anisotropies saturate near the thresholds for the corresponding instability, namely mirror and whistler, respectively.  Moreover, the magnetic moment decreases in time due to pitch angle scattering by the relevant instabilities (Figures \ref{fig:benandanis_wces2500_mime25and64}$e$ and \ref{fig:benandanis_wces2500_mime25and64}$f$).  More quantitatively, the electron pressure anisotropy in simulations with infinite mass ions (where the ions simply provide a neutralizing charge, but do not excite mirror modes) is well described by the linear theory expectation for the whistler instability (see Figures \ref{fig:benandanis_wces1000and5000}$c$ and \ref{fig:benandanis_wces1000and5000}$d$).  For finite mass ratios, however, the electron pressure anisotropy becomes inhomogeneous (Figure \ref{fig:aniso2d}$a$) and $\langle \Delta p_{||,e} \rangle/\langle p_{||,e} \rangle$ is further reduced by a factor of $\sim 1.5-2$ (Figures \ref{fig:benandanis_wces2500_mime25and64}$c$ and \ref{fig:benandanis_wces2500_mime25and64}$d$). We attribute this to the effect of the large-amplitude mirror modes on the electrons, which reduce the growth of the perpendicular electron pressure by bunching the electrons into magnetic mirrors on lengthscales comparable to the ion Larmor radius. The obtained ion pressure anisotropy, $\langle \Delta p_{||,i} \rangle/\langle p_{||,i} \rangle$, is in good agreement with the linear mirror threshold.\newline

\noindent We have also used our simulations to compute the mean free path of particles, $\lambda_{j}$ ($j=i,e$), during the nonlinear stage of the mirror and whistler instabilities.  The average mean free path of both ions and electrons is reasonably well described by
\begin{equation}
\langle \lambda_j \rangle \approx 0.3 \frac{\langle \Delta p_j \rangle}{\langle p_{||,j}\rangle}  \frac{v_{th,j}}{q}
\label{eq:mfp}
\end{equation}
where $q$ ($\equiv s|B_xB_y|/B^2$) is the growth rate of the magnetic field strength, and $s$ is the shear rate (our shear set up is defined by a fluid velocity $\textbf{\textit{v}}=-sx\hat{y}$)\footnote{The shear rate in a turbulent plasma, defined as $k \delta v$, can be dominated by small scales, i.e., high wavenumber $k$.  However, the shear rate that matters here is related to the timescale for the magnitude of $B$ to change, and will thus typically be dominated by large scale dynamics.}.  Physically, this equation describes the balance between pitch-angle scattering by velocity-space instabilities (which limits the pressure anisotropy to $\Delta p_j$) and driving of the pressure anisotropy by the amplification of the background magnetic field at a rate $q$ (see equations \ref{eq:1} \& \ref{eq:2} and associated discussion).\newline  

\noindent Equation \ref{eq:mfp}, together with the relevant instability thresholds, provides a deceptively simple prescription for the ion and electron mean free paths in a low collisionality plasma.  This in turn provides an upper limit on the thermal conductivity of low-collisionality $\beta_j \gtrsim 1$ plasmas.  Of course, these results only apply if the mean free path set by velocity-space instabilities is smaller than the Coulomb mean free path for the plasma under consideration. \newline   

\noindent A second implication of equation \ref{eq:mfp} is that a collisionless plasma has a finite viscosity because the particles do not simply free-stream.  In our simulations with a background velocity shear, the particles are thus heated by tapping into the background shear, just as in a collisional fluid.  We find that the ion and electron heating rates in our simulations are in good agreement with the analytically predicted heating rate by anisotropic viscosity in the limit of a gyrotropic distribution function \citep{SharmaEtAl07}:
\begin{equation}
\frac{d\langle U_j \rangle}{dt} = - s \langle \Delta p_j\rangle \,\frac{B_xB_y}{B^2}
\label{eq:htg}
\end{equation}
The good agreement between equation \ref{eq:htg} and our numerical heating rates in Figure \ref{fig:energygainwhismirro} provides additional support for including this `viscous' heating in models of the thermodynamics of low-collisionality plasmas.  \newline

\noindent The threshold pressure anisotropy found in our simulations is not exactly appropriate for heliospheric and astrophysical plasmas because in the latter the shear rate is much smaller relative to the cyclotron frequency than in our simulations.  The astrophysically relevant threshold for the mirror instability is $\Delta p_i/p_{||,i} \lesssim 1/\beta_i$, while for the electron whistler instability it is 
\begin{equation}
\frac{\Delta p_e}{p_{||,e}} \lesssim \frac{A}{\beta_{||,e}^{0.8}}%   \ \ \ \ \ \ \ \  {\rm whistler}
\label{eq:whistler}
\end{equation}
Equation \ref{eq:whistler} is an approximate fit to the whistler instability threshold relevant for both non-relativistic and relativistic electrons, for growth rates $\gamma_w \sim 10^{-7} \omega_{c,e}$; the relativistic calculations are based on numerical solutions of the dispersion relation derived in \citealt{Gladd1983} (see \citealt{ResslerEtAl15}, Appendix B2).  The  fit is accurate to about 50\% for $\beta_{||,e} \simeq 0.1-100$ (note that \citet{GaryEtAl96} and \citet{SharmaEtAl07} found a somewhat shallower slope $\propto \beta_e^{-0.55}$ in non-relativistic calculations over a smaller range of $\beta_{||,e}$). The coefficient $A$ in equation \ref{eq:whistler} depends weakly on electron temperature, varying from $A \simeq 0.125$ for non-relativistic electrons to $A \simeq 0.25$ for $k_B T_e \simeq 10 m_e c^2$ (relevant to hot accretion flows onto black holes). Finally, in applying equation \ref{eq:whistler} to estimate $\langle \lambda_e \rangle$ (equation \ref{eq:mfp}) and $d\langle U_e\rangle/dt$ (equation \ref{eq:htg}), the reduction in $\langle \Delta p_e \rangle / \langle p_{||,e} \rangle$ by a factor $\sim 1.5-2$ due to nonlinear mirrors should be included.\newline

\noindent  The velocity-space instabilities studied in this paper can impact  the electron pressure anisotropy, mean free path, thermal conduction, and viscous heating in a wide variety of astrophysical environments, including galaxy clusters, low-luminosity accretion flows onto compact objects, and the solar wind.  As a concrete example, we scale our estimate of the electron mean free path to conditions relevant to galaxy clusters using equations \ref{eq:mfp} \& \ref{eq:whistler}:
\begin{equation} \langle \lambda_e \rangle \sim 10 \, {\rm kpc} \, \Big(\frac{f_{M}}{2}\Big)^{-1} \Big(\frac{\beta_e}{100}\Big)^{-0.8} \Big(\frac{T_e}{10^8 \, {\rm K}}\Big)^{1/2} \Big(\frac{q^{-1}}{10^8 \, {\rm yr}}\Big), \label{eq:mfp-clusters} \end{equation}
where $f_{M}$ quantifies the suppression of the thermal conductivity due to mirrors. The precise magnetic growth timescale $q^{-1}$ for clusters is uncertain so we have scaled our estimate to about $0.1$ of the typical cluster dynamical time.   This is likely a modest underestimate at large radii (near the virial radius the timescales are somewhat longer) and a modest overestimate at small radii (in the core the timescales are somewhat shorter).    For comparison, the Coulomb mean free path for a Coulomb logarithm of 10 is
\begin{equation}
\lambda_{C} \approx  0.4  \, {\rm kpc} \, \Big(\frac{T_e}{10^8 {\rm K}}\Big)^2 \Big(\frac{n}{0.1 \, {\rm cm^{-3}}}\Big)^{-1}.
\label{mfp-coul}
\end{equation}
Note that the whistler mediated mean-free path in equation \ref{eq:mfp-clusters} is independent of density.  It may well be shorter than the Coulomb mean free path at large radii in massive (hot) clusters where densities are typically $\sim 10^{-3}-10^{-2} \, {\rm cm^{-3}}$.  This highlights the importance of velocity-space instabilities for understanding the thermodynamics of the outer parts of massive galaxy clusters. \newline

\noindent Our results are also relevant for models of low-luminosity accretion flows onto compact objects.  In particular, our results can be incorporated as sub-grid models for electron conduction and heating in numerical simulations of black hole accretion that attempt to directly predict the emission from the accreting plasma (e.g., \citealt{Monika2014, ResslerEtAl15}).  For concreteness, we note that in the accretion disk context, equation \ref{eq:htg} can be rewritten in terms of the fraction $f_e$ of the total dissipation that goes into the electrons via viscous heating.   The total heating rate per unit volume in magnetized accretion disks is given approximately by $-s B_x B_y/4 \pi$ \citep{BalbusEtAl98}.   As a result, equation \ref{eq:htg} corresponds to
\begin{equation} 
f_e \simeq 0.15 \Big(\frac{f_M}{2}\Big)^{-1} \, \Big(\frac{\beta_e}{100}\Big)^{0.2}, \label{eq:htge} 
\end{equation}
\newline
where we have used equation \ref{eq:whistler}.  Equation \ref{eq:htge} predicts an electron heating rate that is a significant fraction of the total dissipation in accretion disks, and is itself only a weak function of the electron thermodynamics.\footnote{\citealt{SharmaEtAl07} found a somewhat stronger dependence of the electron heating rate on electron temperature, $\propto T_e^{1/2}$.  The difference lies in our treatment of the whistler instability threshold (see equation \ref{eq:whistler} and associated discussion).}\newline

\noindent One limitation of our present study applied to black hole accretion flows is that these environments may be characterized by $T_e \lesssim T_i$, which we have not considered here.  \citet{SironiEtAl2015a} \& \citet{SironiEtAl2015b} argued that for $T_e \ll T_i$ the ion-cyclotron instability becomes more prominent than the mirror instability at $\beta_i \lesssim 100$ and that the electric fields associated with the ion-cyclotron instability can transfer energy directly to the electrons, providing a significant source of electron heating.  The dominance of the ion-cyclotron instability over the mirror instabilty is not captured by the linear stability calculations we have carried out, but for the electrons these are restricted to non-relativistic plasmas.  We also suspect that electron heating by the ion-cyclotron instability is sub-dominant relative to other heating mechanisms (e.g., turbulence and viscous heating) given that for realistic parameters, little of the free energy of the system will reside in the electromagnetic fields associated with the IC waves.  Regardless, however, of this subtlety about the physics of the ion-scale instabilities, the electron pressure anisotropy will still be predominantly regulated by the whistler instability, and so the results presented in this paper are applicable for $T_e \ll T_i$ (aside perhaps from the suppression of the electron mean free path by large amplitude mirrors).

\acknowledgements

We thank Matt Kunz, Lorenzo Sironi, and Alex Schekochihin for useful conversations.  MR thanks the Chilean Comisi\'on Nacional de Investigaci\'on Cient\'ifica y Tecnol\'ogica (CONICYT; Proyecto Fondecyt Iniciaci\'on N$^{\textrm{o}}$ 11121145).  This work was supported by NSF grant AST 13-33612, a Simons Investigator Award to EQ from the Simons Foundation and the David and Lucile Packard Foundation. DV also acknowledges support from NSF/SHINE grant AGS-1460190 and NASA grant NNX16AG81G. We are also grateful to the UC Berkeley-Chile Fund for support for collaborative trips that enabled this work. This work used the Extreme Science and Engineering Discovery Environment (XSEDE), which is supported by National Science Foundation grant number ACI-1053575.

%\clearpage

%\appendix
%\section{Mirror Saturation in 1D and 2D}
%\noindent Previous 1D studies 
\end{document}